\def\Rb{{\Bbb R}}

\def\Cb{{\Bbb C}}

\def\be{\begin{equation}}
\def\ee{\end{equation}}
\def\ba{\begin{array}}
\def\ea{\end{array}}

\documentstyle[11pt]{article}
\input amssym.def
\begin{document}
\baselineskip=18pt \setcounter{page}{1}
\begin{center}
{\LARGE \bf Integrability and PT-symmetry of $N$-Body Systems
          with Spin-coupling $\delta$-Interactions}
\end{center}
\vskip 2mm

\begin{center}
{\normalsize Shao-Ming Fei}
\medskip

\begin{minipage}{4.8in}
{\small \sl
Department of Mathematics, Capital  Normal University, Beijing,
China}\\
{\small \sl Institute of Applied Mathematics, University of
Bonn,  53115 Bonn, Germany}
\end{minipage}
\end{center}

\vskip 2mm
\begin{center}

\begin{minipage}{4.8in}

\centerline{\bf Abstract}
\bigskip

We study the PT-symmetric
boundary conditions for ``spin"-related $\delta$-interactions and
the corresponding integrability
for both bosonic and fermionic many-body systems.
The spectra and bound states are discussed in detail
for spin-$\frac{1}{2}$ particle systems.
\vskip 1mm
PACS number(s): 02.30.Ik, 11.30.Er, 03.65.Fd
\end{minipage}
\end{center}

Integrable models play significant roles in statistical and
condensed matter physics. Many of these models can be exactly
solved in terms of an algebraic or coordinate ``Bethe Ansatz
method" \cite{faddeev}, see e.g., \cite{2345,ladder} for
spin chain and ladder models and \cite{y,y1} for (continuous
variable) quantum mechanical models.
In particular, the quantum mechanical
solvable models describing a particle moving in a local singular
potential concentrated at one or a discrete number of points have
been extensively discussed, see e.g. \cite{agh-kh,gaudin,AKbook}
and references therein. One dimensional problems with point
interactions at, say, the origin ($x=0$) can be characterized by
the boundary conditions imposed on the wave function at $x=0$,
which is equivalent to two particles with contact interactions.
The integrability of one dimensional quantum mechanical many-body
problems with general contact interactions between two particles,
has been studied in \cite{adf} and the bound states and scattering
matrices are calculated for both bosonic and fermionic statistics.
The results are generalized to the case of quantum mechanical
systems with ``spin"-related contact interactions, namely, the
boundary conditions describing the contact interactions are
dependent on the spin states of the particles \cite{afk}.

Recently, the complex generalization of conventional quantum
mechanics has been investigated \cite{bender}. In stead of the
standard formulation of quantum mechanics in terms of Hermitian
Hamiltonians, quantum mechanical models with space-time reflection
symmetry (PT symmetry) have been constructed and studied for
continuous interaction potentials \cite{models}. For point interaction
potentials a systematic description of the boundary conditions and the spectra
properties for self-adjoint, PT-symmetric systems and systems with
real spectra have been presented, and the corresponding integrability of
one dimensional many body systems with these kinds of point
interactions are studied \cite{afkpt}. In this article we study the PT-symmetric
boundary conditions for ``spin"-related $\delta$-interactions and
the corresponding integrability for many-body case.

The family of the usual point interactions for the one dimensional
Schr\"odinger operator $ - \frac{d^2}{dx^2}$
can be described by $2 \times  2$ matrices.
In particular, the boundary conditions describing the
$\delta$-potential interactions have the following form
\begin{equation} \label{bound}
\left( \begin{array}{c}
\varphi\\
\varphi '\end{array} \right)_{0^+}
=\left(
\begin{array}{cc}
1 & 0 \\
c & 1 \end{array} \right)
\left( \begin{array}{c}
\varphi\\
\varphi '\end{array} \right)_{0^-},
\end{equation}
where $c \in \Rb$,
$\varphi(x)$ is the scalar wave function of two spinless particles with
relative coordinate $x$. (\ref{bound}) also describes two particles
with spin $s$ but without any spin coupling between the particles when they
meet (i.e. $x=0$), in this case $\varphi$ represents any one of the
components of the wave function. It is easily verified that the
boundary condition (\ref{bound})
is both self-adjoint and PT-symmetric.

Now consider two particles with spin $s$ and have both $\delta$-interactions
and spin couplings when they meet.
For a particle with spin $s$, the wave function has $n=2s+1$ components.
Therefore two particles with $\delta$-interactions have a general boundary
condition, described in the center of mass coordinate system:
\be\label{BOUND}
\left( \begin{array}{c}
\psi\\
\psi '\end{array} \right)_{0^+}
=\left(
\begin{array}{cc}
{\bf I}_2 & 0 \\
C & {\bf I}_2 \end{array} \right)
\left( \begin{array}{c}
\psi\\
\psi '\end{array} \right)_{0^-},
\end{equation}
where the wave function $\psi$ and its derivative
$\psi '$ are $n^2$-dimensional
column vectors, $C$ is an $n^2\times n^2$ matrix, and
${\bf I}_2$ is the $n^2\times n^2$ identity matrix.

For general $C$ the interactions described by the boundary condition
(\ref{BOUND}) is neither self-adjoint nor PT-symmetric.
From the symmetry condition of the Schr\"odinger operator,
$<-\frac{d^2}{dx^2}u,v>_{L_2(\Rb,\Cb^n)}-
<u,-\frac{d^2}{dx^2}v>_{L_2(\Rb,\Cb^n)}=0$,
for any $u,v\in C^\infty(\Rb \setminus \{0\})$, one has that
(\ref{BOUND}) is self-adjoint when
$C=C^\dagger$, where $\dagger$ stands for the conjugate and transpose.

Applying PT operator to (\ref{BOUND}) we have
$$\left( \begin{array}{c}
\psi\\
-\psi '\end{array} \right)_{0^-}
= \left( \begin{array}{cc}
{\bf I}_2 & 0 \\
C^\ast & {\bf I}_2
\end{array}\right)
\left( \begin{array}{c}
\psi\\
-\psi '\end{array} \right)_{0^+}.
$$
Hence the boundary conditions for the
function $PT \psi$ are given by
$$\left( \begin{array}{c}
\psi\\
\psi '\end{array} \right)_{0^-}
= \left( \begin{array}{cc}
{\bf I}_2 & 0 \\
- C^\ast & {\bf I}_2
\end{array}\right)
\left( \begin{array}{c}
\psi\\
\psi '\end{array} \right)_{0^+},
$$
which coincides with (\ref{BOUND}) if and only if
\begin{equation} \label{sl1} \left( \begin{array}{cc}
{\bf I}_2 & 0 \\
-C^\ast & {\bf I}_2
\end{array}\right) =  \left( \begin{array}{cc}
{\bf I}_2&0\\
C&{\bf I}_2
\end{array} \right)^{-1}.
\end{equation}
This implies that $C^\ast=C$, i.e., $C$ is a real matrix.

Therefore the boundary condition (\ref{BOUND})
is self-adjoint when $C$ is Hermitian
and PT-symmetric when $C$ is real. When $C$ is real symmetric, then
the interaction described by the boundary condition (\ref{BOUND})
is both self-adjoint and PT-symmetric. When $C=c{\bf I}_2$, (\ref{BOUND})
is reduced to the case (\ref{bound}) without spin-coupling
interactions when two particles meet.
In the following we study quantum systems with contact interactions
described by the boundary condition (\ref{BOUND}).
We first consider two spin-$s$ particles case.
The Hamiltonian is then of the form
\be\label{H}
H=(-\frac{\partial^2}{\partial x_1^2}-\frac{\partial^2}{\partial
x_2^2}){\bf I}_2
+2h\delta(x_1-x_2),
\ee
where $h$ is an $n^2\times n^2$ matrix.
If the matrix $h$ is proportional to the unit matrix
${\bf I}_{2}$, $H$ is reduced to the usual
two-particle Hamiltonian with contact interactions but no spin
coupling.

Let $e_\alpha$, $\alpha=1,...,n$,
be the basis (column) vector with the $\alpha$-th component as $1$
and the rest components $0$. The wave function of the system (\ref{H})
is of the form
\be\label{psi}
\psi=\sum_{\alpha,\beta=1}^n\phi_{\alpha
\beta}(x_1,x_2)e_\alpha\otimes e_\beta.
\ee
In the center of mass coordinate system, $X=(x_1+x_2)/2$, $x=x_1-x_2$,
the operator (\ref{H}) has the form
\begin{equation}
H = - \left( \frac{1}{2} \frac{\partial^2}{\partial X^2}
+ 2 \frac{\partial ^2}{\partial x^2} \right)
{\bf I}_{2} + 2 h \delta (x).
\end{equation}
The functions $\phi = \phi (x,X)$ from the domain of this operator
satisfy the following boundary conditions at $x=0$,
\be\label{b2}
\ba{l}
\phi_{\alpha\beta} '(0^+,X)-\phi_{\alpha\beta} '(0^-,X)=
\sum_{\alpha,\beta=1}^n h_{\gamma\lambda,\alpha\beta}
\phi_{\gamma\lambda}(0,X),\\[3mm]
\phi_{\alpha\beta}(0^+,X)=\phi_{\alpha\beta}(0^-,X),~~~\alpha,\beta=1,...,n,
\ea
\ee
where the indices of the matrix $h$ are arranged as $11,12,...,1n$; $21,22,...,
2n$; ...; $n1,n2,...,nn$. (\ref{b2}) is just the boundary condition (\ref{BOUND})
with $C=h$. Here $h$ acts on the basis vectors of the two particles
as $h \,(e_\alpha\otimes e_\beta) =\displaystyle
\sum_{\gamma,\lambda=1}^n h_{\alpha\beta,\gamma\lambda} e_\gamma\otimes
e_\lambda$.

According to the statistics
$\psi$ is symmetric (resp. antisymmetric) under the interchange of
the two particles if $s$ is an integer (resp. half integer).
Let $k_1$ and $k_2$ be the momenta of the two particles.
In the region $x_1<x_2$, in terms of Bethe hypothesis the
wave function has the following form
\be\label{w1}
\psi=u_{12}e^{i(k_1x_1+k_2x_2)}+u_{21}e^{i(k_2x_1+k_1x_2)},
\ee
where $u_{12}$ and $u_{21}$ are $n^2\times 1$ column matrices.
In the region $x_1>x_2$,
\be\label{w2}
\psi=(P^{12}u_{12})e^{i(k_1x_2+k_2x_1)}
+(P^{12}u_{21})e^{i(k_2x_2+k_1x_1)},
\ee
where according to the symmetry or antisymmetry conditions,
$P^{12}=p^{12}$ for bosons and $P^{12}=-p^{12}$ for fermions, $p^{12}$
being the operator on the $n^2\times 1$ column that interchanges
the spins of the two particles.
Substituting (\ref{w1}) and (\ref{w2}) into the boundary
conditions (\ref{b2}), we get
\be\label{a1}
\left\{
\begin{array}{l}
u_{12}+u_{21}
=P^{12}(u_{12}+u_{21}),\\
ik_{12}(u_{21}-u_{12})
=hP^{12}(u_{12}+u_{21})+ik_{12}P^{12}(u_{12}-u_{21}),
\end{array}\right.
\ee
where $k_{12}=(k_1-k_2)/2$. Eliminating the term $P^{12}u_{12}$ from
(\ref{a1}) we obtain the relation
\be\label{2112}
u_{21} = Y_{21}^{12} u_{12}~,
\ee
where
\be\label{a21a12}
Y_{21}^{12}=[2ik_{12}{\bf I}_{2} -h]^{-1}[2ik_{12}P^{12}+h].
\ee

For a system of $N$ identical particles with $\delta$-interactions,
the Hamiltonian is given by
\be\label{HN}
H=-\sum_{i=1}^N\frac{\partial^2}{\partial x_i^2}{\bf I}_N+\sum_{i<j}^N h_{ij}
\delta(x_i-x_j),
\ee
where ${\bf I}_N$ is the $n^N\times n^N$ identity matrix,
$h_{ij}$ is an operator acting on the $i$-th and $j$-th bases as $h$ and the
rest as identity, e.g., $h_{12}=h\otimes {\bf 1}_3\otimes...{\bf 1}_N$,
with ${\bf 1}_i$ the $n\times n$ identity matrix acting on the $i$-th particle.
The wave function in a given region, say $x_1<x_2<...<x_N$, is of the form
\be\label{Psi}
\ba{rcl}
\Psi&=&\displaystyle\sum_{\alpha_1,...,\alpha_N=1}^n
\phi_{\alpha_1,...,\alpha_N}(x_1,...,x_N)
e_{\alpha_1}\otimes...\otimes e_{\alpha_N}\\[4mm]
&=&u_{12...N}e^{i(k_1x_1+k_2x_2+...+k_Nx_N)}
+u_{21...N}e^{i(k_2x_1+k_1x_2+...+k_Nx_N)}\\[3mm]
&&+(N!-2)\, {\rm other~terms},
\ea
\ee
where $k_j$, $j=1,...,N$, are the momentum of the $j$-th particle.
$u$ are $n^N\times 1$ matrices.
The wave functions in the other regions are determined
from (\ref{Psi}) by the requirement of
symmetry (for bosons) or antisymmetry (for fermions).
Along any plane $x_i=x_{i+1}$, $i\in 1,2,...,N-1$, we have
\be\label{a1n}
u_{\alpha_1\alpha_2...\alpha_j\alpha_{j+1}...\alpha_N}=
Y_{\alpha_{j+1}\alpha_j}^{jj+1}
u_{\alpha_1\alpha_2...\alpha_{j+1}\alpha_j...\alpha_N},
\ee
where
\be\label{y}
Y_{\alpha_{j+1}\alpha_j}^{jj+1}=[2ik_{\alpha_j\alpha_{j+1}}{\bf I}_N-h_{j{j+1}}]^{-1}
[2ik_{\alpha_j\alpha_{j+1}}P^{jj+1} + h_{jj+1}].
\ee
Here $k_{\alpha_j\alpha_{j+1}}=(k_{\alpha_j}-k_{\alpha_{j+1}})/2$ play
the role of momenta
and $P^{jj+1}=p^{jj+1}$ for bosons and $P^{jj+1}=-p^{jj+1}$ for fermions,
where $p^{jj+1}$ is the operator on the $n^N\times 1$ column $u$
that interchanges the spins of particles $j$ and $j+1$.

For consistency the operators $Y$ in (\ref{y})
must satisfy the Yang-Baxter equation with
spectral parameter \cite{y,y1},
$$
Y^{m,m+1}_{ij}Y^{m+1,m+2}_{kj}Y^{m,m+1}_{ki}
=Y^{m+1,m+2}_{ki}Y^{m,m+1}_{kj}Y^{m+1,m+2}_{ij},
$$
or
$$
Y^{mr}_{ij}Y^{rs}_{kj}Y^{mr}_{ki}
=Y^{rs}_{ki}Y^{mr}_{kj}Y^{rs}_{ij}
$$
if $m,r,s$ are all unequal, and
\be\label{ybe2}
Y^{mr}_{ij}Y^{mr}_{ji}=1,~~~~~~
Y^{mr}_{ij}Y^{sq}_{kl}=Y^{sq}_{kl}Y^{mr}_{ij}
\ee
if $m,r,s,q$ are all unequal.
By a straightforward  calculation it can be shown that the operator $Y$ given
by (\ref{y}) satisfies all the Yang-Baxter relations if
\be\label{hp}
[h_{ij}, P^{ij}]=0.
\ee
Therefore if the Hamiltonian operators for the spin coupling commute with the
spin permutation operator, the $N$-body quantum system (\ref{HN}) can be
exactly solved. The wave function is then given by (\ref{Psi}) and
(\ref{a1n}) with the energy $E=\sum_{i=1}^N k_i^2$.

For the case of spin-$1\over 2$, $h$ is a $4\times 4$ matrix. A
matrix satisfying (\ref{hp}) is generally of the form
\be\label{hspin}
h=\left(\ba{cccc}
a&e_1&e_1&c\\
e_2&f&g&e_3\\
e_2&g&f&e_3\\
d&e_4&e_4&b\ea\right),
\ee
where $a,b,c,d,f,g,e_1,e_2,e_3,e_4\in\Cb$.
When $e_2=e_1^\ast$,  $e_4=e_3^\ast$,
$d=c^\ast$ and $a,b,f,g\in\Rb$, the interaction is reduced to the
self-adjoint case \cite{afk}. when $a$, $b$, $c$, $d$, $f$, $g$, $e_1$, $e_2$, $e_3$,
$e_4\in\Rb$, the interaction is PT-symmetric, and when
$e_2=e_1$,  $e_4=e_3$, $d=c$ and $a,b,c,f,g,e_1,e_3\in\Rb$,
the interaction is both self-adjoint and PT-symmetric.

One can compare this kind of quantum mechanical integrable systems with
the integrable spin chain models.
Let $V$ denote a complex vector space. A matrix $R$
taking values in $End_c(V\otimes V)$ is called a solution of the quantum
Yang-Baxter equation (QYBE) without spectral parameters, if it satisfies
${\cal R}_{12}{\cal R}_{13}{\cal R}_{23}=
{\cal R}_{23}{\cal R}_{13}{\cal R}_{12}$,
where ${\cal R}_{ij}$ denotes the matrix on the complex vector space
$V\otimes V\otimes V$, acting as $R$ on the $i$-th and the $j$-th
components and as identity on the other components. When $V$ is two
dimensional, the solutions of QYBE include
the ones of the form (\ref{hspin}) \cite{ybe}, although generally
(\ref{hspin}) is not a solution of the QYBE. A solution $R$ in principle
corresponds to a kind of spin coupling operator between the nearest
neighbor spins and gives rise to an integrable spin chain model
by Baxterization \cite{ma}.
Therefore for an $N$-body system to be integrable,
the spin coupling in the contact interaction (\ref{hspin}) is allowed to be
more general than the spin coupling in a Heisenberg spin chain model with
nearest neighbors interactions.

The spectra of (\ref{HN}) with $h$ given by (\ref{hspin}) are generally
complex. They are real when the spectra of $h$ are real. The self-adjoint
case is included in the parameter family that gives real spectra. The
real spectra related to the PT-symmetric case are divided into two parts:
self-adjoint and non-Hermitian.

When $h$ gives purely real spectra,
The many-body system (\ref{HN}) possesses also bound states.
For $N=2$, from (\ref{a1}) the bound states have the form,
\be\label{bpsi2}
\psi^2_\alpha=u_\alpha e^{\frac{c+a\Lambda_\alpha}{2}\vert x_2-
x_1\vert},~~~~\alpha=1,...,n^2,
\ee
where $u_\alpha$ is the common $\alpha$-th eigenvector of the operators
$h$ and $P^{12}$, with eigenvalue $\Lambda_\alpha$,
such that $hu_\alpha=\Lambda_\alpha u_\alpha$ and $c+a\Lambda_\alpha<0$,
$P^{12}u_\alpha=u_\alpha$.
The eigenvalue of the Hamitonian $H$ corresponding to the bound state
(\ref{bpsi2})
is $-(c+a\Lambda_\alpha)^2/2$. In stead of a unique bound state
for $\delta$-interaction with boundary condition (\ref{bound}),
here we have $n^2$ bound states. For the $N$-particle system
the bound states are then given by
\be\label{bpsin}
\psi^N_\alpha=v_\alpha e^{-\frac{c+a\Lambda_\alpha}{2}\sum_{i>j}\vert
x_i-x_j\vert},~~~\alpha=1,...,n^2,
\ee
where $v_\alpha$ is the wave function of the spin part.
It can be checked that $\psi^N_\alpha$ satisfy the boundary condition
(\ref{b2}) at $x_i=x_j$ for any $i\neq j\in 1,...,N$. The spin wave
function $v$ here satisfies $P^{ij}v_\alpha=v_\alpha$ and
$h_{ij}v_\alpha=\Lambda_\alpha v_\alpha$, for any $i\neq j$.
The energy of the bound state $\psi^N_\alpha$ is
$E_\alpha=-(c+a\Lambda_\alpha)^2N(N^2-1)/12$.
The corresponding scattering matrices can be similarly studied
for real spectra case.

We have studied the boundary conditions with spin-coupling
$\delta$-interactions including self-adjoint, PT-symmetric and real spectrum
families. The corresponding integrability
for both bosonic and fermionic many-body systems
are investigated. Detailed examples are given to the case of spin-$\frac{1}{2}$
particle systems, where the spectra and bound states are also discussed.
The results can be straightforwardly generalized to higher spin particle systems.
The contact interaction we considered here is just $\delta$-interaction.
For general contact interactions with spin-couplings, the PT-symmetries,
spectra problems, and integrability of many-body problem
could be much more complicated and remain to be investigated further.

\end{document}